\documentclass[usenatbib]{aastex}
\shorttitle{Comet Family Transit}
\shortauthors{Bodman \& Quillen}

\begin{document}
\title{KIC 8462852: Transit of a Large Comet Family}
\author{Eva H. L. Bodman, Alice Quillen}
\affil{Department of Physics and Astronomy, University of Rochester}
\affil{Rochester, NY 14627, USA}
\email{ebodman@pas.rochester.edu}

\begin{abstract}
We investigate the plausibility of a cometary source of the unusual transits observed in the KIC 8462852 light curve. A single comet of similar size to those in our solar system produces a transit depth of the order of $10^{-3}$ lasting less than a day which is much smaller and shorter than the largest dip observed ($\sim20\%$ for $\sim3$ days), but a large, closely traveling cluster of comets can fit the observed depths and durations. We find that a series of large comet swarms, with all but one on the same orbit, provides a good fit for the KIC 8462852 data during Quarters 16 and 17, but does not explain the large dip observed during Quarter 8. However, the transit dips only loosely constrain the orbits and can be fit by swarms with periastrons differing by a factor of 10. To reach a transit depth of $\sim0.2$, the comets need to be in a close group of $\sim30$, if they are $\sim100$ km in radius or in a group of $\sim300$ if they are $\sim10$ km in radius. The total number of comets required to fit all of the dips is $\sim70$ $\sim$100 km or $\sim700$ $\sim10$ km comets. A single comet family from a tidally disrupted Ceres-sized progenitor or the start of a Late Heavy Bombardment period explains the last $\sim60$ days of the unusual KIC 8462852 light curve.
\end{abstract}
\keywords{comets: general--- stars: individual (KIC 8462852)--- stars: peculiar}

\section{Introduction}

\citep{2015B} reported the discovery of KIC 8462852, a twelfth magnitude star in the \textit{Kepler} field \citep{2010B}, by the Planet Hunter project \citep[e.g.][]{2012F}. The light curve displays two unusual dip events: a smooth 15\% dip around day 800 of the main \textit{Kepler} mission and a complex series of irregular dips starting around day 1500, the largest of which is $\sim20$\%. Using optical spectra, \cite{2015B} find the star to be an ordinary F3V star and find no excess infrared (IR) emission from \textit{WISE} data taken before either dip. The lack of excess has been confirmed with recent IR measurements \citep{2015L,2015M,2015T}. The lack of IR excess greatly constrains the total mass of warm dust, but a large amount of dust could have been created and dissipated between the \textit{WISE} observation and the recent measurements. \cite{2015W} speculated that alien megastructures could be a possible cause. \cite{2015B} propose a family of comets as the most probable source for the unusual transit dips. We investigate this hypothesis here for the day 1500 event.

\cite{1996B} proposed that a family of sungrazing comets could explain the transient spectral features observed in the $\beta$ Pictoris system. Referred to as the falling evaporating body (FEB) model, dust from icy bodies falling into the star sublimates causing temporary absorption lines. These features last for longer than the transit time of a single comet, but a family of comets in which each comet reaches periastron a couple hours after the previous one can account for the apparent long duration \citep{1996B}. The comet families likely formed from a disruption of one comet or multiple bodies trapped in mean motion resonance \citep{2014K}.  Expanding from the FEB model, \cite{1999LV} simulated photometric variations from a single comet dust tail occulting the star. Their model predicts a $10^{-4}-10^{-2}$ dip in flux that depends mostly on the comet's size and periastron distance. For most orbits, dust tail transits have a characteristic ``rounded triangle'' shape in the light curve from a sharp drop in flux followed by a gradual increase, but for a small range of longitude of periastron, they are approximately symmetric in time.

Expanding upon the comet transit model by \cite{1999L}, we develop a simple outgassing comet model that includes sublimation and multiple comet nuclei to fit the complex dips observed in the KIC 8462852 light curve data from Quarters 16 and 17. We do not attempt to fit the first large dip at day 800 as its shape, a gradual decrease in flux followed by a sharp increase, is the opposite of a typical comet transit light curve and not well explained by a simple model. Implausibly well-timed comets would be necessary to fit the curve. In this work, we assume this earlier dip is unrelated to the day 1500 dips. 

\section{Model}

We develop a comet family transit model by expanding the single comet transit model by \cite{1999L}. First, we track the dust distribution over a single comet orbit with a particle simulation, and then use copies of the dust distribution, shifted in time to represent comets on the same orbit reaching periastron at different times, to form a family. With the dust distribution from the entire family, we calculate the time-dependent total dust extinction in the line of sight to compute variations in flux.

\begin{figure}
\includegraphics[width=6.0in, trim= 0 0 0 0 ]{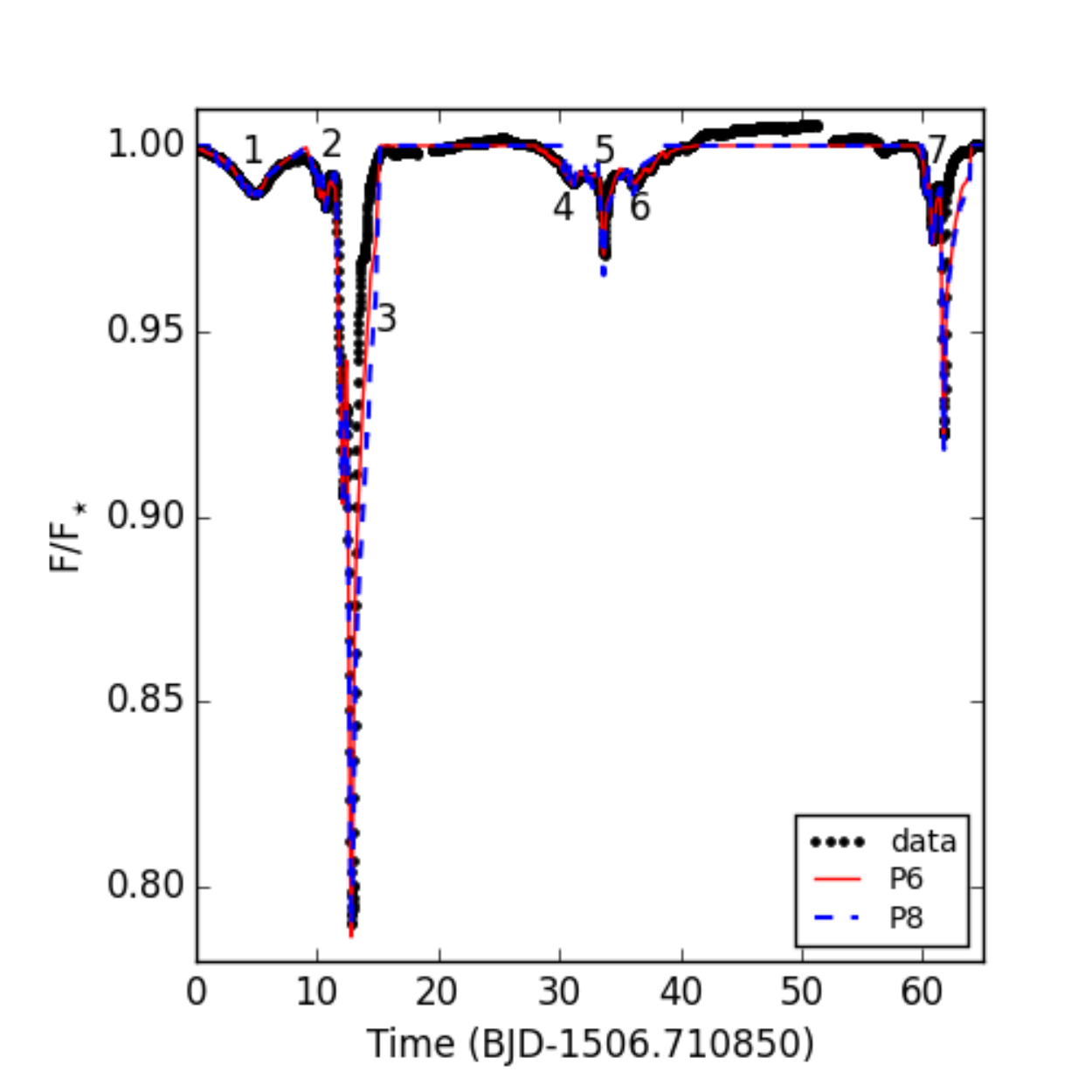}
\caption{Normalized \textit{Kepler} data are black dots with dips labeled for reference. The P8 (100 km) and P6 (10 km) comet family fits are the dashed blue and solid red lines, respectively. The P6 fit is smoother but has a larger number of comets, see Table \ref{tab:families}.
}
\label{fig:fit}
\end{figure}

We use the normalized PDCSAP{\_}FLUX data for KIC 8462852 from the \textit{Kepler} mission during Quarters 16 and 17 when the irregular complex of dips occurs and we do not attempt to fit any of the other earlier dips. This portion of the light curve is shown in Fig. \ref{fig:fit} as dots, and, for convenience, we labeled the dips from 1 to 7. We did not attempt to subtract the 0.88 day stellar rotation cycle or the 10-20 day cycle found by \cite{2015B} since those fluctuations ($\lesssim0.1\%$) are small compared to the dips. 

Our model uses comet orbital parameter inputs: periastron within range $0.05<q<1$ AU, longitude of periastron ($\varpi$) and inclination ($i$) with respect to the line of sight and the argument of periastron ($\omega$). The other input parameter is dust production rate, which is a proxy for the comet nucleus size.  Assuming simple dust production from sublimation of H$_2$O ice releasing dust \citep[e.g.,][]{2002I}, the dust production rate is proportional to the area, $P/(1\ \mathrm{kg\ s}^{-1})\approx 10^4(R/1\ \mathrm{km})^2$. We set the apastron to 20 AU for all simulations so that the orbital period ($\sim$30 years) is much longer than the \textit{Kepler} observing period and set the stellar mass, temperature, and luminosity to a F3V star \citep{2015B}. To keep the model simple, we restricted each dip to a group of comets on identical orbits with the same production rate except for a small variation in inclination (see Sec. \ref{sec:results}). We also varied the time between each comet individually.

\subsection{Dust Distribution}

The dust distribution is set by the radiation pressure, the grain size distribution, the dust production rate, and sublimation. Particles representing many dust grains are produced isotropically from a point source (the comet nucleus) with a 1 km s$^{-1}$ ejection velocity and then travel along a trajectory defined by gravity and radiation pressure to form the comet tail. The particles are propagated along Keplerian orbits, but with an effectively reduced gravitational field compared to the comet nucleus, $g_\mathrm{eff}=GM_\star(1-\beta)/r^2$. Here, $M_\star$ is the mass of the star, $G$ is the gravitational constant, $r$ is the distance between the particle and the star, and $\beta$ is the ratio of radiation pressure to gravitational force. Following \cite{1999LV}, we define the ratio as
 \begin{equation}
 \beta =0.2 \left({L_\star/M_\star\over L_\odot/ M_\odot}\right)\left({s\over 1 \mu\mathrm{m}}\right)^{-1}
 \end{equation}
where $L_\star$ is the stellar luminosity and $s$ is the radius of the dust grain. We assume a power law grain size distribution with an exponential small particle cut-off proposed by \cite{2012FR}, $dn/ds\propto (s_p\alpha/ s)^\alpha\mathrm{exp}(-s_p\alpha/s)$. Here, the constant $s_p$ is the peak of the distribution and $\alpha$ is a constant. We choose typical values of $\alpha=4$ and $s_p=0.5$ \citep{2012FR}. We set the lower and upper radius limits at 0.01 and 10.0 $\mu$m.

Following \cite{1999L}, we adopt a distance-dependent dust production rate,
\begin{equation}
P(r)=P_0\left({r_0\over r}\right)^2\left(L_\star\over L_\odot\right)
\end{equation}
where the characteristic distance is $r_0=1$ AU and $P_0$ is the characteristic dust production rate, which also sets comet size. The dust production is taken to be zero beyond a distance of $r_\mathrm{crit}=3\sqrt{L_\star/L_\odot}$ AU.

We expanded the model developed by \cite{1999L} to include a simple dust sublimation model. Our sublimation model assumes spherical grains of a single chemical composition that reach thermal equilibrium on a timescale that is small compared to the evaporative lifetime, see \cite{2014V}. If the particle temperature is above 1000K, we compute the temperature-dependent rate of change in that particle's size and remove it from the simulation if $s<0.01 \ \mu \mathrm{m}$. Our results are insensitive to the sublimation rate.

We calculated the dust distribution for a single comet using a time step of $\sim0.5$ hr. To model a comet family, we used copies of that single comet dust distribution, but each is offset by its own constant time delay.

\subsection{Extinction}

\begin{figure}
\includegraphics[width=6.0in, trim= 0 0 0 0 ]{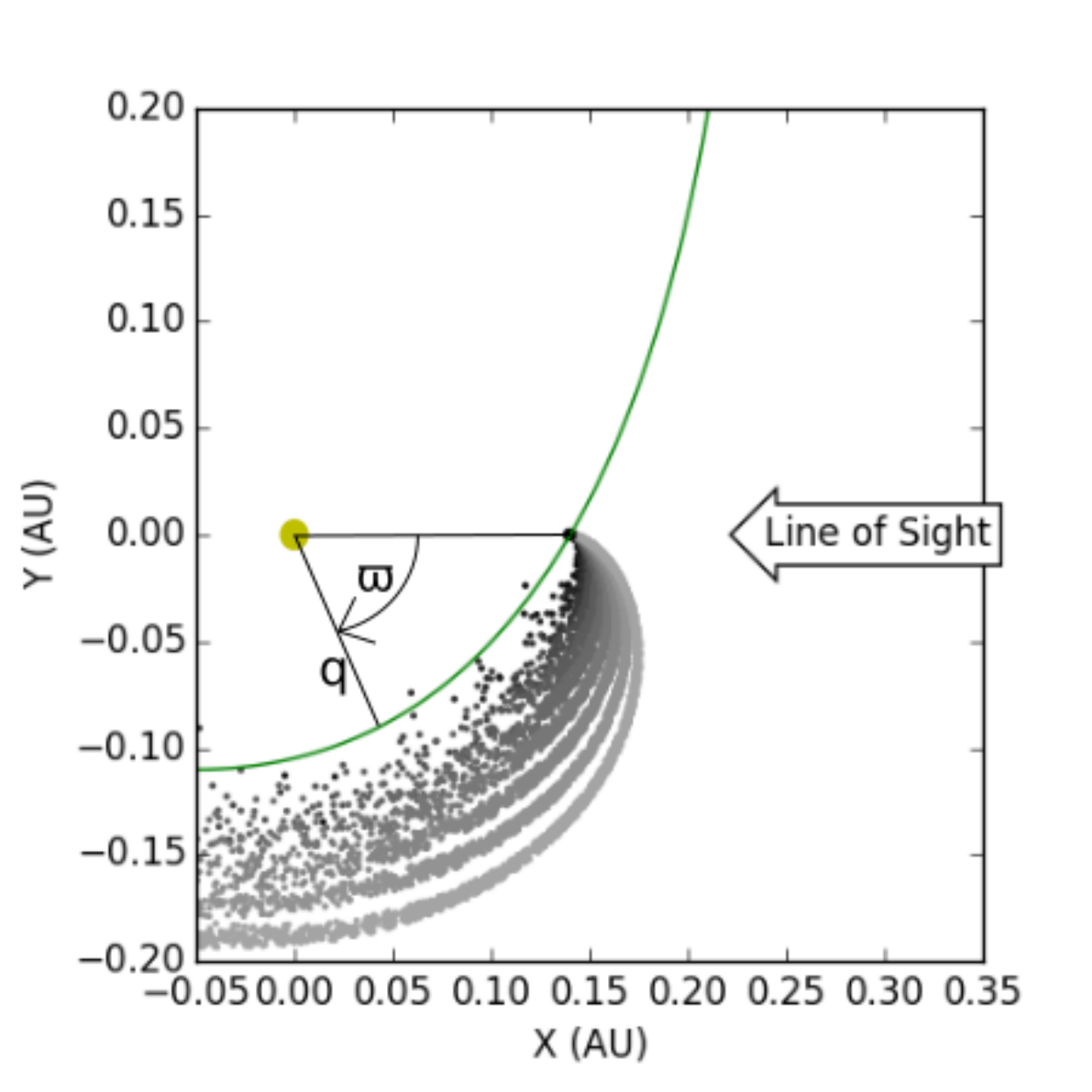}
\caption{Illustration of a single comet with the longitude of the periastron ($\varpi\approx-60^\circ$) labeled and the comet orbit in green. The black and gray dots are the comet nucleus and dust tail particles. The star is the yellow dot. The tail's feathery appearance is a numerical artifact.
}
\label{fig:comet}
\end{figure}

We model the light curve by calculating the time-dependent dust extinction from the comet tails occulting the star. Following \cite{1999LV}, we calculate the total dust extinction by adding up the extinction from each particle in the line of sight to the star, see Fig. \ref{fig:comet}. The optical depth of the dust is
\begin{equation}
\tau={\Sigma_\mathrm{part.}N_\mathrm{grain/part}Q_\mathrm{ext}(s,\lambda)\pi s^2\over S}
\end{equation}
where $S$ is the area of the star projected into the line of sight, $Q_\mathrm{ext}$ is the extinction efficiency, and $N_\mathrm{grain/part}$ is the number of dust grains per model particle. The total number of particles in a single comet simulation is $\sim20,000$.  The extinction cross section of a dust grain is $Q_\mathrm{ext}\pi s^2$. Using the approximation made by \cite{1999L} for dust grains larger than 0.1 $\mu$m, we set $Q_\mathrm{ext}(s,\lambda)=(s/\lambda)^4+s/\lambda$ if $s<\lambda$ and set $Q_\mathrm{ext}=2$ if $s\geq\lambda$.

Ignoring limb darkening, we divide the area of the star into a grid of cells with area $S_i$ where $\Sigma_iS_i=\pi R_\star^2$ and then calculate the optical depth of each cell. The ratio of the flux observed through the dust to the initial stellar flux is
\begin{equation}
{F_\mathrm{ext}\over F_\star}=\sum_i {S_ie^{-\tau_i}\over\pi R_\star^2}.
\end{equation}
Using the time-dependent dust distribution, we calculate the stellar flux as a function of time.

\section{Results}\label{sec:results}

To investigate the hypothesis of a comet family origin of the KIC 8462852 light curve dips, we attempted to model the all of the dips occurring after day 1500 with a single orbit. We varied the dust production rate for each comet group, but could not find a satisfactory fit for dip 1 on the same orbit as the rest. Our fits use $\varpi=-80^\circ$ for all dips, but $q=0.2$ for dip 1 and $q=0.1$ AU for the others. Most comets had no inclination, but dips 3, 5, and 7 required some on slightly inclined orbits. Table \ref{tab:families} summarizes the comet properties of each group for two comet dust production rate criteria, $P_0\leq 10^8$ kg s$^{-1}$ (P8) and $P_0\leq 10^6$ kg s$^{-1}$ (P6). The fits for each are shown with blue (dashed) and red (solid) lines in Fig \ref{fig:fit}, respectively. The number and timing of the comets varied greatly with production rate, 73 $\sim100$ km radius comets several hours apart for P8, and 731 $\sim10$ km comets every couple of hours for P6. The dip durations and depths are reasonably matched with the noticeable exceptions being rate of flux increase after dips 3 and 7 where the model rate is slightly too slow, more so in the P8 fit than the P6. The slopes are also not as smooth as the data for many of the dips in the P8 fit.

We made two fits with different dust production criteria because there is degeneracy between the production rate and the number of comets. A reduction in production rate can be compensated by an increase in the number of comets. We used the $\beta$ Pic system to guide our choice. To match the observed transient features in the spectra of the $\beta$ Pic system, \cite{1996B} assume a dust production rate of $P_0\approx 10^6\ \mathrm{kg\ s}^{-1}$ which is $10^3$ times higher than for typical solar system bodies and estimate a peak activity rate of $\sim10$ comets per day. In order to reach a depth of 0.2 with a similar comet rate, $P_0=10^{8}$ kg s$^{-1}$ is required, so we used that for our first fit. For that production rate, the comets are large, of the order of 100 km in radius. When we reduce the dust production rate to that of the $\beta$ Pic system for our second fit, the number of comets increases by a factor of 10 but the comets are only of the order of 10 km. The smaller size restriction also decreases the time delay between comets, which is a constraint on any disruption model for the comet family and the dynamical history.

We chose the family orbit that minimized the number of comets needed to fit the largest dip (3) and then increase periastron for dip 1 by a small amount to achieve a good fit. For dip 1, the duration was too long ($\sim10$ days) to fit without greatly increasing the number of comets or using comets too large to produce a smooth dip. We used the criterion of minimizing the number of comets because the periastron is not well constrained from the slope or duration of the dips. For each comet group, good fits exist for periastrons varying by up to a factor of 10. The dip 3 group is most constrained in periastron, which is within a factor of two of our model.  While a single comet family does fit the data, the comet groups are not restricted to a single orbit and the data can be fit with multiple families. The reason for the uncertainty is the use of multiple bodies. For a single object of a given size, the slope and transit duration are set by the speed of the object so a distance can be estimated. However, for multiple comet bodies, the slope and transit duration can be fitted to some extent by adjusting the time delay. In Table \ref{tab:families}, we report the average time interval over the entire dip, but the actual time between comets varies significantly. For example, the first three comets of dip 7 in the P8 fit are much more than five hours apart to match the slower decrease and small increases in flux, but the last five are a half hour apart. For the P6 fit, the number of comets is much larger and the spacing is much smaller, but, otherwise, the variance in timing is similar.

Since each individual dip is somewhat symmetrical in time, the longitude of periastron was constrained to less than -75$^\circ$ because the comet tail is more aligned with the line of sight when traveling away from the star. For an F3 star, radiation pressure is not strong enough to orient the dust tail straight out radially so, instead, the densest region of the tail mostly follows the nucleus along most of the orbit. After periastron, the direction of the tail deviates more from the orbit and can become approximately aligned with line of sight, causing the flux dip to be more symmetric. At $\varpi=-80^\circ$, the tail being misaligned with the line of sight is the source of the discrepancy between model and data in dips 3 and 7, but decreasing $\varpi$ does not improve the fit.

There is no constraint on the inclination of the orbits from the dips, so an inclination of zero was assumed except for some comets in dips 3 and 7 in the P8 fit and dips 3, 5, and 7 in P6. If the orbit is inclined, the comet transit depth decreases and duration shortens, but these effects can be masked by adjusting the number of comets. For dip 3, the 20\% drop in flux requires a large fraction of the star's area to be occulted by dust. Despite being less than half the depth, dip 7 and dip 5 also require very closely orbiting comets to match the slopes. If all comets are at the same inclination, the transit depth does not increase with increasing number of comets or a higher dust production rate. Neither adjustment increases the area occulted by the dust since the comet nuclei are too close together and their dusty tails overlap. In order to reach those transit depths on those timescales, we set some of the comets in that family to a non-zero inclination and the argument of periastron $\omega=-20^\circ$. For the P8 fit, we used only one inclination, $i=1^\circ$, and for the P6 fit, used a range of inclinations $i=\pm1^\circ,\pm2^\circ$. Neither of these values is constrained by the fit, but a constraint on the time since the comet disrupted would limit $\omega$ because the disruption would cause the spread in $i$.

\section{Discussion}

The large number of comets necessary to fit the KIC 8462852 light curve produces a significant amount of warm dust. The entire comet family produces $\sim10^{-7}$ $M_\oplus$ of dust, about 70\% of which is created during dip 3. Assuming a single temperature for the dip 3 group dust near periastron, the fractional flux of the dust in the near-IR and mid-IR are $\sim10^{-2}$ and $10^{-1}$; thus, during such a transit, IR excess is detectable with space-based observations such as \textit{WISE}, but quickly diminishes as the dust disperses. The dust mass is within limits of recent far-IR measurements \citep{2015T}. Temporary gas signatures are also likely detectable, such as those found for some young A stars \citep[e.g.][]{2012M,2014K}.

Sungrazing comets are produced with a planet by three different mechanisms: the Kozai mechanism, secular resonance, and mean motion resonance. \cite{1992B} showed that the Kozai mechanism can efficiently produce sungrazers from highly inclined bodies in the asteroid belt. The Kozai mechanism is rotationally invariant with no preferential periastron direction; thus, to produce a family on similar orbits, an object needs to be disrupted. The $\nu_6$ secular resonance has also been shown to put icy bodies in the asteroid belt on sungrazing orbits \citep{1994F}. Secular resonances require at least two planets and are sensitive to planet configuration. So while $\nu_6$ is strong in the solar system, other secular resonances could have a similar affect for other configurations and produce an asymmetry in the periastron direction that put multiple bodies on similar orbits \citep{1994L}. If a planet is migrating through a disk, planetesimals are efficiently trapped in mean motion resonance where they can reach sungrazing orbits \citep[e.g.][]{2000Q,2001T} that are asymmetric in the periastron direction. This mechanism is also insensitive to planetary configuration.

Assuming the comets are composed of ice (1 g cm$^{-3}$), a 100 km radius comet has a mass of $4\times10^{21}$ g and a 10 km comet is $4\times10^{18}$ g. The total mass of the comets is 10$^{-5}\ M_\oplus$ for the P8 fit and 10$^{-7}\ M_\oplus$ for P6. For the Kozai mechanism, this would require a $R\approx400$ km icy body on a highly inclined orbit to be disrupted and cross our line of sight, which would be a rare event. The total comet mass is consistent with resonance sweeping through a disk during late migration, such as predicted by the Nice Model \citep{2010M}.  All three mechanisms can produce the large comet family to cause the KIC 8462852 light curve dips. However, mean motion resonance is the most probable because the mechanism is general and produces several progenitors on similar orbits instead of relying on one single very large icy body disruption.

Stellar perturbations can cause comet showers \citep{1981H}. \cite{2015B} found a possible M-dwarf companion 2" away that may have been a stellar perturber for a comet shower, but its orbit and closest approach time and distance are unknown. Stellar perturbations cause asymmetry in the periastron direction that is necessary to produce a cluster of transit dips, but the star's approach needs to be very close to produce a large shower \citep{2005D}. A companion 1000 AU away might be able to produce the number of comets needed, but comets would be on many different orbits.

\section{Conclusion}

We find that a comet family on a single orbit with dense swarms can cause most of the observed complex series of irregular dips that occur after day 1500 in the KIC 8462852 light curve. However, the fit requires a large number of comets and is also not well constrained. We cannot limit the structure of the system much beyond the observational constraints and the dynamical history of the comet family is unknown, but, if the comet family model is correct, there is likely a planetary companion forming sungrazers. Since the comets are still tightly clustered within each dip, a disruption event likely occurred recently within orbit, such as tidal disruption by the star. This comet family model does not readily explain the large dip observed around day 800 and treats it as unrelated to the ones starting at day 1500. The flux changes too smoothly and too slowly to be easily fit with a simple comet family model, but a swarm with comets that transit in gradually increasing frequency and suddenly ends can adequately fit the data. If the comets are large ($\sim100$ km) for the day 800 dip, the same swarm can cause both dipping events.

\acknowledgments
We thank Tabetha Boyajian for introducing us to this fascinating light curve and Eric Mamajek for useful comments. This research has made use of the NASA Exoplanet Archive, which is operated by the California Institute of Technology, under contract with the National Aeronautics and Space Administration under the Exoplanet Exploration Program. This work was supported in part by NASA grant NNX13AI27G.

\begin{deluxetable}{lcccccccccccc}
\tablecolumns{9}
\tablewidth{0pc}
\tablecaption{Comet Groups}
\tablehead{
\colhead{}  & \colhead{}   &  \multicolumn{3}{c}{P8 ($R_\mathrm{comet}\lesssim$100 km)} &\colhead{}  & \multicolumn{3}{c}{P6 ($R_\mathrm{comet}\lesssim$10 km)} \\
\cline{3-5} \cline{7-9} \\
\colhead{Group} & \colhead{$q$ (AU)}   & \colhead{$P_0$ (kg s$^{-1}$)}    & \colhead{$\Delta t$(hr)} & \colhead{$N_\mathrm{comets}$}   &\colhead{}& \colhead{$P_0$}    & \colhead{$\Delta t$} & \colhead{$N_\mathrm{comets}$}}

\startdata
1         & 0.2   &    10$^7$      &   9      &  12        &&   10$^6$   & 5     & 65          \\
2         & 0.1   &    10$^7$      &   7      &  5          &&   10$^6$   & 3     & 27          \\
3         & 0.1   &    10$^8$      &   0.9   &  36/10  &&   10$^6$   & 0.2 & 441/341  \\
4         & 0.1   &    10$^7$      &   12    &  5          &&   10$^6$   & 8     & 37           \\
5         & 0.1   &    10$^8$      &   3      &  3          &&   10$^6$   & 0.7  & 24/8      \\
6         & 0.1   &    10$^7$      &   7      &  3          &&   10$^6$   & 8     & 24           \\
7         & 0.1   &    10$^8$      &   5      &  9 /3      &&   10$^6$   & 1     & 113/60    \\
\sidehead{Total Comets:}
           &        &                       &           &    73     &&                 &   & 731 \\
\enddata

\tablecomments{The dip associated with each group is labeled in Fig. \ref{fig:fit}. The time between comets, $\Delta t$, is averaged. In the $N_\mathrm{comet}$ columns, the total number of comets in the group is reported, then the number of inclined comets, if any. The longitude of periastron for all comets is $\varpi=-80^\circ$.
\label{tab:families}}
\end{deluxetable}

\end{document}